\documentclass[12pt,a4paper,titlepage]{article}
\usepackage{amsmath,amsthm,amsfonts}
\usepackage[active]{srcltx}
\usepackage{graphicx}
\begin{document}

\newcommand{\EF}{\mathcal{E}}
\newcommand{\abs}[1]{\left\lvert #1 \right\rvert}
\newcommand{\norm}[1]{\left\lVert #1 \right\rVert}
\newcommand{\ket}[1]{\lvert #1 \rangle}
\newcommand{\brkt}[2]{\langle #1 \vert #2 \rangle}
\newcommand{\braket}[3]{\langle #1\lvert #2\rvert #3\rangle}
\newcommand{\ketbra}[2]{\lvert #1 \rangle\langle #2 \rvert}
\newcommand{\schrod}{Schr\"odinger }
\newcommand{\bs}{\negthickspace}
\newtheorem{lemma}{Lemma}
\newtheorem*{thm}{Theorem}
\newtheorem{unbounded}{Theorem}
\parskip = 0.5\baselineskip

\begin{titlepage}
\begin{center}
{ \tt }

 {\huge \bf Gravitational Repulsion}

 \vspace{0.1cm} {\huge \bf within a Black-Hole}

 \vspace{0.1cm} {\huge \bf using the Stueckelberg}

 \vspace{0.1cm} {\huge \bf  Quantum Formalism}

{ \tt }

 {\large \bf }\vspace{0.5cm}

 { \tt Dedicated to the memory of a great physicist:

  }
 { \tt \large \bf  John Archibald Wheeler}\vspace{0.3cm}

{ \tt }

 {\large \bf }

 { \tt by}\vspace{0.1cm}

 {\large \bf D. M. Ludwin$^1$ and L. P. Horwitz$^2$}

 \vspace{0.1cm}
 {\small $^1$Department of Physics, Technion, Haifa 32000, Israel}

 \vspace{0.02cm}
 {\small $^2$School of Physics, University of Tel Aviv, Ramat Aviv 69978, Israel}

\end{center}
\textbf{Abstract}\\
We wish to study an application of Stueckelberg's relativistic quantum theory in the framework of
general relativity. We study the form of the wave equation of a massive body in the presence of a Schwarzschild gravitational field. We treat the mathematical behavior of the wavefunction also around and beyond the horizon ($r=2M$). Classically, within the horizon, the time component of the metric becomes spacelike and distance from the origin singularity becomes timelike, suggesting an inevitable propagation of all matter within the horizon to a total collapse at $r=0$. However, the quantum description of the wave function provides a different understanding of the behavior of matter within the horizon. We find that a test particle can almost never be found at the origin and is more probable to be found at the horizon. Matter outside the horizon has a very small wave length and therefore interference effects can be found only on a very small atomic scale. However, within the horizon, matter becomes totally "tachionic" and is potentially "spread" over all space. Small location uncertainties on the atomic scale become large around the horizon, and different mass components of the wave function can therefore interfere on a stellar scale. This interference phenomenon, where the probability of finding matter decreases as a function of the distance from the horizon, appears as an effective gravitational repulsion.

\end{titlepage}
\pagenumbering{arabic}
\section{Introduction}

In Newton's classical mechanics and in quantum mechanics, one makes use of a global time that has causal meaning. In standard non-relativistic quantum mechanics, \emph{time} is interpreted as a causal parameter, where, for each value of the parameter, the quantum states are coherent. The manifestly covariant quantum Stueckelberg formalism is based on the idea that there is an invariant parameter $\tau$ of evolution of the system; wave functions, as covariant functions of space and the Einstein time $t$ form a Hilbert space (over $R^4$) for each value of $\tau$. Thus, there are two types of time, one transforming covariantly, and the second, a parameter of evolution \cite{HAE89}.

J. A. Wheeler was one of the first physicists to understand the physical difference between both kinds of time. Together with his student, Richard P. Feynman, they showed the significance of both the retarded and the advanced parts of the particle propagator for the correct description of direct particle interaction \cite{WF49} (in the 5D electrodynamics associated with the Stueckelberg-Schr\"odinger equation, the $\tau$-retarded propagator contains both $t$-retarded and advanced components \cite{AH09}). The symmetry between past and future in the prescription of the fields as a consequence of the QED theory, leads to the understanding that a causal parameter in our experience, i.e., an invariant universal \emph{time}, may not be totally correlated with the Minkowski or Einstein time, in which interactions notice past and future simultaneously. This means that the system admits quantum superposition of states in the total 4D picture where time acts as another spatial dimension and events can move forward and backward in time $t$. Later, with his famous \emph{Delayed Choice} thought experiment \cite{WH83}, Wheeler showed that a measurement may intervene in the static 4D picture and alter the whole 4D picture in its entirety. The intervening measurement not only changes the "present" state (an incident that would disturb unitary evolution and other conservation laws), but actually transforms the entire "history" of the 4D state and therefore maintains unitarity; four-momentum is conserved \cite{LB01}.

In relativity, the time of an event as measured in the laboratory is subject to variation according to the velocity of the apparatus related to the transmitting system and may as well be affected by forces (such as gravity). A relativistic quantum theory should therefore incorporate \emph{time} in a manifestly covariant manner and also permit definition of a global \emph{causal} parameter to generate evolution of the 4D states.

To describe the dynamical evolution of such a system, Stueckelberg and Horwitz-Piron \cite{ST41,Larry73} introduced an invariant parameter $\tau$, a "world time", coinciding with the Newtonian world time, accounting for classical as well as quantum relativistic evolution. The world time $\tau$ provides a parameter that labels the dynamical evolution of the covariant system. For free on shell motion of a single particle, the Einstein proper time can be taken equal to the world time.

We shall make use of this Manifestly Covariant Quantum Theory (MCQT) introduced by Stueckelberg in 1941 and generalized by Horwitz and Piron \cite{Larry73} in 1973. The theory has the structure of Hamilton dynamics with the Euclidean three dimensional space replaced by four-dimensional Minkowski space. Since all four components of energy-momentum are kinematically independent, the theory is intrinsically "off-shell".

The self-interaction problem of the relativistic charged particle has recently been studied, where it was shown that the radiation field is associated with excursions from the mass shell \cite{OH01,OH01_2,KOH04,OH06,RH09}. In the limit that the motion maintains a state very close to the mass shell limit, the equation reduces \cite{OH01} to that of Dirac \cite{Dir38} (the Abraham-Dirac-Lorentz equation). The recent experiment of Lindner et al \cite{Lin05}, showed the phenomenon of interference in time, an essential property of quantum mechanical wave functions coherent in time, as in the Stueckelberg theory. The results were predicted by Horwitz and Rabin in 1976 \cite{Larry51}, and the significance of this important experiment was discussed in \cite{Larry241}.

Another result which concerns the Stueckelberg MCQT formalism which was pointed out recently \cite{Larry228}, is that by using the metric tensor in the kinematical terms of the Stueckelberg-Schrodinger equation one can obtain classical general relativity in eikonal approximation. Since the eikonal approximation lowers the dimension of the differential equations describing the fields by one \cite{Visser01}, the eikonal approximation to the five dimensional Stueckelberg quantum equation in a curved spacetime, characterized by a metric tensor $g_{\mu\nu}$, results in the four dimensional Einstein geodesic equations.

When dealing with a gravitational potential represented by the metric tensor within the covariant theory, the theory remains on-shell and can be shown to have a form similar to the Klein-Gordon (KG) equation on a curved spacetime for each value of the mass (on a continuum).

In this paper we study an application of Stueckelberg's MCQT in general relativity. We deal with a simple case and compare the results to those expected in classical general relativity. We study the form of the wave equation of a test particle in the presence of a Schwarzschild gravitational field, assuming that the source is massive enough to ignore changes in the metric caused by the smaller mass of the test particle. J. Makela \cite{Makela2002} has considered the black hole as an atom in the framework of quantum theory, but did not discuss the spatial distribution of the wave functions. Since the Stueckelberg formalism describes the evolution of the wave function of the body according to the invariant world time of evolution, we can treat the mathematical behavior of the wavefunction also around and beyond the horizon ($r=2M$). Within the horizon, the interval classically effectively changes its signature and becomes spacelike; the distance from the origin singularity becomes timelike, but the description according to world time enables us to study the behavior of the evolution of the wavefunction in an an absolute sense.

We find that within the black hole horizon, the expectation value of the distance from the $R=0$ singularity has a strong gradient towards the horizon. This result can only be explained quantum mechanically, since classically, the particle should move towards the origin. Interference effects evidently induce results which are very different from the classical behavior. The interference makes a difference in the results only when the particle has a wavelength of stellar scale, $\lambda \sim M$. In the neighborhood of the horizon, the wavefunction is evidently spread, in analogy to the action of tidal forces. This result means that for certain wave-packets, there is "gravitational repulsion" that prevents the test particle from falling towards the $R=0$ singularity, and maintains it near the horizon in a manner that partly depends on the particle's angular momentum. Therefore, this quantum gravitational model predicts that the test particle  will move to the $r=2M$ shell of the black hole, which means that matter in the black hole should accumulate on the interior of the horizon. As we have indicated above, in the eikonal (ray) approximation, the quantum solutions flow as in the classical limit, and our interpretations are therefore a valid description of the expected observed behavior of a particle as a result of the application of the quantum theory.

The dynamics associated with the matter repulsion within the black hole, causes the Einstein time to run in the opposite direction of the world time. This effect is well-known in the Stueckelberg-Feynman diagram of electron-positron annihilation, where the positron is interpreted as an electron going "backwards" in time. This effect around the horizon will cause matter within the black-hole to have properties of antimatter. This model may explain some part of the matter-anti-matter asymmetry observed in the universe, if there are sufficient black hole surfaces. Using covariant quantum theory, therefore, may help provide a new understanding of black hole physics and cosmology.

This paper is in memory of John Archibald Wheeler, who passed away the week this paper had started being written. In this paper we combine some of his ideas regarding time, together with the physics of the "black hole", a term which he came up with and brought to worldwide attention.

\section{The Eikonal Approximation}\label{section eikonal approximation}

We start with the Stueckelberg-Schrodinger (Horwitz-Oron) equation \cite{Larry233,Larry234}:

\begin{equation}\label{HOS}
    i\frac{\partial\Psi}{\partial\tau}=\frac{1}{2m\sqrt{g}}\partial^\mu
    g_{\mu\nu}\sqrt{g}\partial^\nu \Psi
\end{equation}
where $m$ is an intrinsic property of the particle with dimension
mass.

The Stueckelberg Hamiltonian is written with a non-trivial metric type function $g_{\mu\nu}$ on spacetime in the quadratic kinetic term.

We will show hereafter, that for the eikonal (semi-classical) approximation, these equations lead to Einstein's geodesic flow on a curved manifold; general relativity then appears as an emergent phenomenon \cite{Larry233}.

In the eikonal approximation for the 5D equation above, we assume a $\Psi$ such that:

\begin{equation}
    \Psi(x,\tau)=A(x)e^{-i(\frac{\kappa}{2m}\tau-\frac{\sqrt{\kappa}}{m}S(x))}
\end{equation}

which gives:
\begin{equation}
    \frac{\kappa}{2m} A(x)e^{-i(\frac{\kappa}{2m}\tau-\frac{\sqrt{\kappa}}{m}S(x))}=-\frac{\kappa}{2m}
    g_{\mu\nu}\frac{\partial^{\mu}S\partial^{\nu}S}{m^2}A(x)e^{-i(\frac{\kappa}{2m}\tau-\frac{\sqrt{\kappa}}{m}S(x))}
    + O (\sqrt{\kappa}) + \dots
\end{equation}
 In the eikonal approximation, it is assumed that $\kappa$ ($\kappa=p_\mu p^\mu\equiv -\tilde{m}^2$, the dynamical measured mass squared; we use the signature (-,+,+,+) in the local flat space) is large compared to the square of the second derivative of $S$; the dynamical evolution of the system in $\tau$ is effectively frozen, and the theory reduces to a four dimensional eikonal form:

\begin{equation}
    g_{\mu\nu}\frac{\partial^{\mu}S\partial^{\nu}S}{m^2}=-1
\end{equation}

In the optical analogy of the eikonal approximation
\cite{KlineKay}, the functions $S(x)$ are the Fresnel surfaces of rays, and $\partial^{\mu}S=p^\mu$ is the momentum in the direction of the
propagation of the phase surface. If we identify the momentum $p^\mu$ as $m{\dot{x}}^\mu$ where the dot corresponds to differentiation in $\tau$, this equation becomes:
\begin{equation}\label{eq. Einstein invariant}
    -g_{\mu\nu}dx^\mu dx^\nu=d\tau^2
\end{equation}
the equation for the invariant line element of Einstein.

The Schr\"{o}dinger current associated with (\ref{HOS}) is \cite{KlineKay,MTW}:
\begin{equation}
    j_\tau(x)_\nu=\frac{1}{2mi}(\Psi_\tau^*g_{\mu\nu}\partial^\mu\Psi_\tau-\Psi_\tau g_{\mu\nu}\partial^\mu\Psi_\tau^*)
\end{equation}

The Fresnel surface of the system's dynamics, analogous to the optical case is \cite{KlineKay}:

\begin{equation}
    K=g_{\mu\nu}p^{\mu}p^{\nu}+m^2=0
\end{equation}

It is clear that $\partial K/\partial p_\mu$ is in the direction of the eikonal form of $j_\tau^\mu$. This implies that $K$ is the evolution operator for the dynamical flow of the particles which correspond to the Fresnel rays. $K$ is therefore, the covariant Hamiltonian
of the system. It then follows from the Hamilton equations that the flow is geodesic, where $g_{\mu\nu}$ is the metric \cite{MTW}.

We note that according to the Stueckelberg theory, $ds/d\tau=\tilde{m}/m$, where $\tilde{m}^2=g_{\mu\nu}p^\mu p^\nu$, and this change of variable in (\ref{eq. Einstein invariant})
brings the line element to the form using proper time $ds$ with a factor $\tilde{m}/m$, said to be "on shell" if this quantity is unity.

\section{The Stueckelberg Formalism in a Schwarzschild
gravitational field}

For the Schwarzschild metric we have (we consider the purely radial case for which $\phi = 0$, and take G=1):
\begin{equation}
g^{\mu \nu }=\left(
\begin{array}{llll}
 \frac{1}{1-\frac{2 M}{r}} & 0 & 0 & 0 \\
 0 & \frac{2 M}{r}-1 & 0 & 0 \\
 0 & 0 & -\frac{1}{r^2} & 0 \\
 0 & 0 & 0 & -\frac{\csc ^2 \theta}{r^2}
\end{array}
\right)
\end{equation}
and also:
\begin{equation}
    \sqrt{-g}=r^2 \sin \theta
\end{equation}
After substituting the metric into equation (\ref{HOS}) and separating variables, we find that the solution to the problem is of the form:
\begin{equation}
    \Psi _{\tau }\left(x^{\mu }\right)=e^{-i k   \tau -i \omega t-i \alpha \varphi }
   P_{l,\alpha }(\theta ) R_{\kappa ,\omega ,l}(r)
\end{equation}
where $\kappa=2mk$ has the dimensions of $mass^2$, and equals $m^2$ on the particle's mass shell.

Substituting this form into equation (\ref{HOS}), we get an equation for $R_{\kappa ,\omega ,l}(r)$:

\begin{equation}\label{R equation}
R''(r)+\frac{2 (r-M)}{r
   (r-2 M)} R'(r)+\left(\frac{r^2 \omega ^2}{(r-2 M)^2}-\frac{\kappa  r^2+l
   (l+1)}{r (r-2 M)}\right) R(r)=0
\end{equation}

and for $P_{l,\alpha }(\theta )$:

\begin{equation}\label{P equation}
P''(\theta)+\cot{\theta} P'(\theta )+\left(l (l+1)-\frac{\alpha ^2}{\sin ^2 \theta}
   \right) P(\theta )=0
\end{equation}

\subsection{The form of the gravitational potential in the weak gravity limit ($r>>2M$)}
For the far gravitational field, where we take $r>>2M$ and therefore neglect high orders of $\frac{2M}{r}$, after substituting $\kappa\rightarrow m^2$, the radial equation has the form:

\begin{equation}\label{eq. radial far}
R''(r)+\frac{2}{r} R'(r)+(\omega^2-m^2-\frac{l(l+1)}{r^2}+\frac{2M(2\omega^2-m^2)}{r}) R(r)=0
\end{equation}

If we divide the whole equation by $-2m$ we find that this equation has the exact form of the Schrodinger central potential equation, where the term $\frac{l(l+1)}{2m r^2}$ represents the repelling \emph{centrifugal potential}. The Centrifugal term shall be omitted hereafter, since we shall keep only the first order terms of $\frac{1}{r}$.

Since, relativistically, $\omega$ is the energy of the particle including its mass and since we are not interested here in tachyonic solutions, we take $\omega^2- m^2$ to be positive. Therefore, as opposed to the nonrelativistic hydrogen problem, the eigenvalues of interest are positive. The tachyonic solutions exponentially decrease at infinity and therefore don't add any relevant amplitude to the far field solutions; however, one could argue that they should be taken into consideration near the horizon.

The effective central potential for this equation has the form:

\begin{equation}\label{eq. far field Sh gravitation potential}
    V(r)=-\frac{M(2\omega^2-m^2)}{m r}\approx -\frac{M}{r}(m+\frac{2\textbf{p}^2}{m})\approx -\frac{mM}{r}(1+2v^2)
\end{equation}

The effective \emph{gravitational mass} is altered by a small factor of $2v^2$. Moving back to ordinary units, the effective \emph{gravitational mass} is: $\mu=m(1+\frac{2 v^2}{c^2})$.

The solution to the radial equation is of the form:
\begin{equation}\label{eq. Solution to far radial}
    R(r)=\frac{\text{c1} J_1\left(2 \sqrt{\mu M r }\right)+\text{c2}
   Y_1\left(2 \sqrt{\mu M r }\right)}{\sqrt{r}}
\end{equation}

where $J_n(z)$ and $Y_n(z)$, give the Bessel function of the first and second kind accordingly.

The fact that the energy eigenvalues of the equation are positive means that the wavefunctions, as single momentum modes, are not square integrable.
We consider therefore wave packets which include many momentum modes. For the far fields, our massive object may be localized very tightly, and may have an uncertainty in its location which is very small compared to the scale of the gravitational distances. This situation, where the object's wavelength doesn't play a role in the physics of the problem, changes when the metric causes the wavelength (or equivalently the location uncertainty) of the object to be of relevant scale, which will cause interference phenomena to be crucial.

We now discuss predictions close to and within the black hole horizon, and in a following section show the results of a complete and exact computation of the wavefunction in the entire interior region.

\subsection{Solutions for the radial part in special cases}
Taking:
\begin{equation}\label{R2B tranform equation}
R(r)=\frac{B(r)}{\sqrt{r} \sqrt{r-2 M}}
\end{equation}
and substituting in equation (\ref{R equation}), we get for $B_{m ,\omega ,l}(r)$ a Schrodinger-like equation of the form:

\begin{equation}\label{B equation}
B''(r)-\frac{r(r-2M) \left(l(l+1)+m^2 r^2\right)-M^2-\omega ^2 r^4} {(r-2M)^2 r^2}B(r)=0
\end{equation}

\subsubsection{Solving the radial equation when approaching the horizon}

A general solution of the radial equation seems to be very difficult to achieve. Therefore, we solve the equation for those regions where simplifying assumptions permit an easier calculation. We know that such assumptions contribute to a certain amount of inaccuracy, and therefore we can treat the results only qualitatively and not quantitatively. The solutions give us a good idea of the direction in which the body will move. When the body moves to other regions, we must take into account the other parts of the potential, and the calculations become more complicated.

We first study the behavior of particles near the horizon ($r\rightarrow 2M^+$).
Much work has been done on the behavior of the field equations when approaching the Schwarzschild horizon from above, for example, the most general solution of the radial equation in an asymptotic form, under the assumption of independency of time (which is a good assumption at the horizon), is given in \cite{CF77}.

Expanding the potential around $r\rightarrow2M^+$, the potential in equation (\ref{B equation}) takes the form:

\begin{equation}\label{potential at horizon}
   U(d)= -\frac{\omega ^2}{d^2}-\frac{2 \omega ^2-m^2}{d}+\frac{l (l+1)+\frac{1}{2}}{4
   M^2 d}
\end{equation}
where $d=\frac{r-2M}{2M}\rightarrow0$.

This is clearly a very strong attracting potential (apart from the relatively neglected weak repelling centrifugal part).

Solving the radial equation in this region we obtain for the dominant parts:

\begin{equation}\label{wave equation at horizon}
    R(d)=a_1 d^{-2 i M \omega
   }+a_2 d^{2 i M \omega
   }
\end{equation}

Finding the expectation value of $r\rightarrow2M^{+}$, with normalization, we need to compute:

\begin{equation}
    <r>=\frac{\int_{2 M}^{2 M+\lambda } 4\pi  r^3 R(r) R(r)^* \, dr}{\int_{2
   M}^{2 M+\lambda } 4 \pi  r^2 R(r) R(r)^* \, dr}
\end{equation}

    where $\lambda$ is the uncertainty in the location of the particle on the $R$ axis or, equivalently, the width of the wave packet associated with the wavelength of the particle.

Changing variables around the horizon we get:

$$ <r>=\frac{M \left(a_1
   a_2 \left((3 \epsilon
   -4 i M \omega ) \epsilon
   ^{8 i M \omega }+4 i M
   \omega \right)+16 M^2
   \omega ^2 \epsilon ^{4 i M
   \omega }\right)}{a_1
   a_2 \left((\epsilon
   -2 i M \omega ) \epsilon
   ^{8 i M \omega }+2 i M
   \omega \right)+8 M^2 \omega
   ^2 \epsilon ^{4 i M \omega
   }}\rightarrow 2M $$

   where $\epsilon=\frac{\lambda}{2M}<<1$.

What can be seen from the result is the fact that the particle is very strongly captured by the horizon. The uncertainty in its location is now related to the phase in the term $\epsilon ^{i M \omega}$ which becomes exponentially small and concentrated near the horizon through normalization of the wavefunction.

\subsubsection{Solving the radial equation at $r\rightarrow2M^{-}$}

Taking the same potential from equation (\ref{potential at horizon}) and solving the radial equation for $r\rightarrow2M^{-}$ we obtain (only the first term of (\ref{wave equation at horizon}) applies to the interior solution):

\begin{equation}
    R(d)=a_1 d^{-2 i M \omega}
\end{equation}
where $d=\frac{2M-r}{2M}\rightarrow0$.

Finding the expectation value of $r$ in a small neighborhood of the horizon, i.e., at $r=2M-\rho$, where $\rho<<2M$ we need to compute:

\begin{equation}
    <r>=\frac{\int_{2 M-\rho-\lambda}^{2 M-\rho+\lambda } 4\pi  r^3 R(r) R(r)^* \, dr}{\int_{2 M-\rho-\lambda}^{2 M-\rho+\lambda } 4 \pi  r^2 R(r) R(r)^* \, dr}
\end{equation}

Changing variables around the horizon we get:

$$ <r>=2 M-\rho +\frac{2 \lambda ^2}{6 M-3 \rho }$$

It can be seen that within a small distance from the horizon at the point $r=2M-\rho$, the expectation value tends towards $r=2M$.
This result means that effectively, a particle that is at the horizon, will most probably stay there and not fall into the center.

Using $\lambda$ in the expectation value calculation indicates a width for computing the local probability. Since we are using $\lambda$ as the effective spread of the particle's wave function (and beyond that we have a zero probability of finding the particle) the calculation is a good approximation. Of course, if $\lambda$ is spread beyond the validity of our approximation, this assumption isn't justified.

The reason that the interacting particle behaves differently quantum-mechanically than is expected classically, is the fact that the expectation value takes interference into account. A similar phenomena is the reason for which the electron in its ground state doesn't fall into the atom. One can think of this as the result of the quantum effect of interference (see also \cite{Makela2002}).

\subsubsection{Solving the radial equation when approaching r=0}

The equations are well defined within the horizon. Therefore, it is of interest to study the equation when $r\rightarrow0$.

When $r\rightarrow0$ $(r<<M)$, the potential in equation (\ref{B equation}) takes the form (a repulsive effective potential):

\begin{equation}\label{potential at origin}
   U(r)= -\frac{2 l (l+1)+1}{4 M r}-\frac{1}{4 r^2}
\end{equation}

We get for the radial equation the solution:

\begin{equation}\label{eq. wavefunction at origin}
    R(r)=a_1+a_2 \log \left(\frac{r}{M}\right)
\end{equation}

Finding the expectation value of $r$ very close to the origin, at $r=\rho$, where $\rho<<2M$ we need to compute:

\begin{equation}\label{result at origin}
    <r>=\frac{\int_{\rho-\lambda}^{\rho+\lambda } 4\pi  r^3 R(r) R(r)^* \, dr}{\int_{\rho-\lambda}^{\rho+\lambda } 4 \pi  r^2 R(r) R(r)^* \, dr}
\end{equation}

after some series expansion we get:

$$ <r>\approx\rho+\frac{2 \lambda ^2 \rho }{\lambda ^2+3 \rho ^2}$$

\begin{figure}
\begin{center}
  \includegraphics{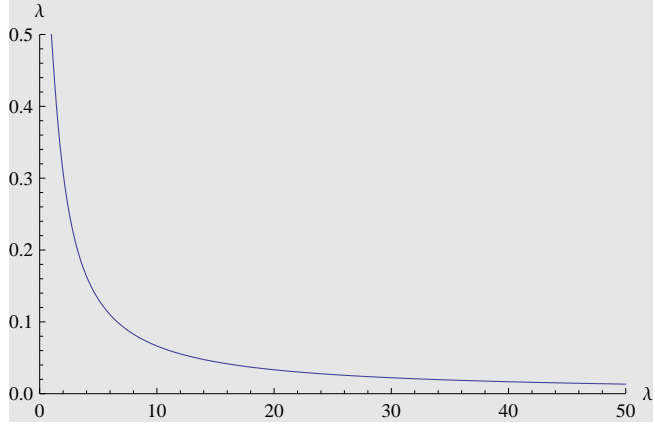}\\
  \end{center}
  \caption{The amount of "repulsion" from the particle's location (in $\lambda$ units) as function of the distance from $r=0$ (in $\lambda$ units)}\label{repulsion}

\end{figure}

The point we want to show in (\ref{result at origin}) is, that the expectation value for a "particle" located initially at $r=\rho$ is greater than $\rho$. Of course the expectation value is still small and doesn't "explode" at $r=0$, and also the "repulsion" (i.e., the distance between the location $\rho$ and the expectation value) becomes smaller and smaller as you draw away from $r=0$ (Figure \ref{repulsion}); however, this is a consistent result, and will cause matter eventually not to be found close to the origin even if it starts there. Therefore we can summarize, that, although the potential shows a very strong attractive force, the quantum mechanical equation implies a repulsive behavior.

When $\rho\rightarrow\lambda$, becomes smaller (if for instance the black hole was created with mass around the origin), the effect of repulsion becomes even more dominant. At the limit where $\rho<\lambda$, the expectation value becomes:

$$ <r>= \frac{\int_{0}^{\rho+\lambda } 4\pi  r^3 R(r) R(r)^* \, dr}{\int_{0}^{\rho+\lambda } 4 \pi  r^2 R(r) R(r)^* \, dr}
\approx\frac{3 (\lambda +\rho )}{4} $$

The result we obtain shows that mass effectively moves from within the \emph{black hole} to the \emph{horizon}.

\subsection{Exact computation of the wavefunction in the entire interior region}
In the previous section, we have computed approximations to the wavefunction in three short ranges of radius, one near the origin, one near the horizon from within and a third near the horizon from without. However, in order to validate our conclusions from the previous section, we shall perform a calculation with a single and globally defined wavefunction that allows comparison of the probability residing in separate regions. We shall, therefore, solve (\ref{R equation}) numerically for particular selected values of $\kappa$ and $\omega$ and with proper boundary conditions, and give a full picture of the wave function in the whole (internal) region.

\subsubsection{Constructing the initial wavefunction}

For the numerical evaluation we shall choose: $M=100$. Our particle's energy shall be described by a gaussian distribution function around a main energy of $\omega_0=2.5$ with a frequency width of $\sigma_\omega=0.2$. The mass of the particle shall be taken to be constant (with no distribution) over the whole wavefunction in order that the numerics fit the KG equation for a specific mass; $m=\sqrt{\kappa}=\frac{\omega_0}{2}=1.25$.
We shall also take the trivial case where $l=0$, and so the wave function is independent of $\theta$.

Since $t$ is spacelike within the interior, our initial conditions for the wavefunction are a well localized gaussian in $t$ (and in $\omega$) with width $\sigma=M/10=10$ ($\sigma_\omega=0.2$ in the frequency domain). The gaussian is constructed numerically using $100$ eigenfunctions with different frequencies around the central frequency $\omega_0$, with a width of $\Delta_\omega=1$,  such that the minimal frequency $\omega_0-\Delta_\omega$, is still greater than $m$ and therefore we avoid tachionic modes in the solution.

The wavefunction, as function of the spatial coordinates $t$ and $r$, is therefore described by the superposition of the states according to:
\begin{equation}\label{Numerical Gaussian Construction}
    \Phi(r,t)=\sum_{i=1..100}R_{\omega_i}(r)e^{-i \omega_i t}
\end{equation}

\begin{figure}
\begin{center}
  \includegraphics{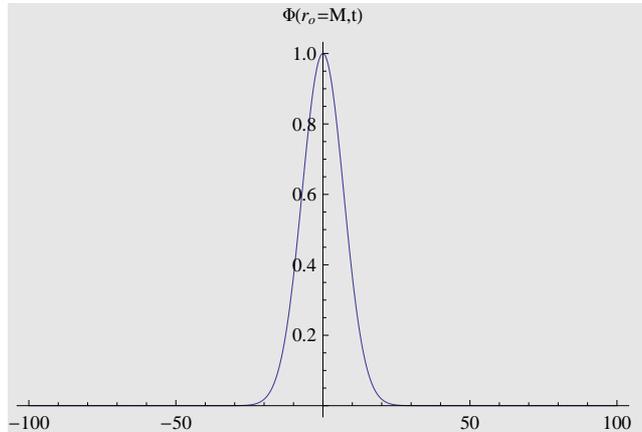}\\
  \end{center}
  \caption{The boundary condition for the wavefunction at $r_0=M$}\label{initGaussian}
\end{figure}

The first boundary condition on the evolution of $\Phi(r,t)$ is the gaussian as function of $t$, shown in Figure \ref{initGaussian}, when taking all the $100$ frequency modes at $r_0=M$.

The second boundary condition on the evolution of states shall be $R'_{\omega_i}(r_0=M)=0$.

\subsubsection{The "time-like" evolution of the wavepacket}
We shall now let the function evolve in 'r', which is thought as the timelike axis.

\begin{figure}
\begin{center}
  \includegraphics{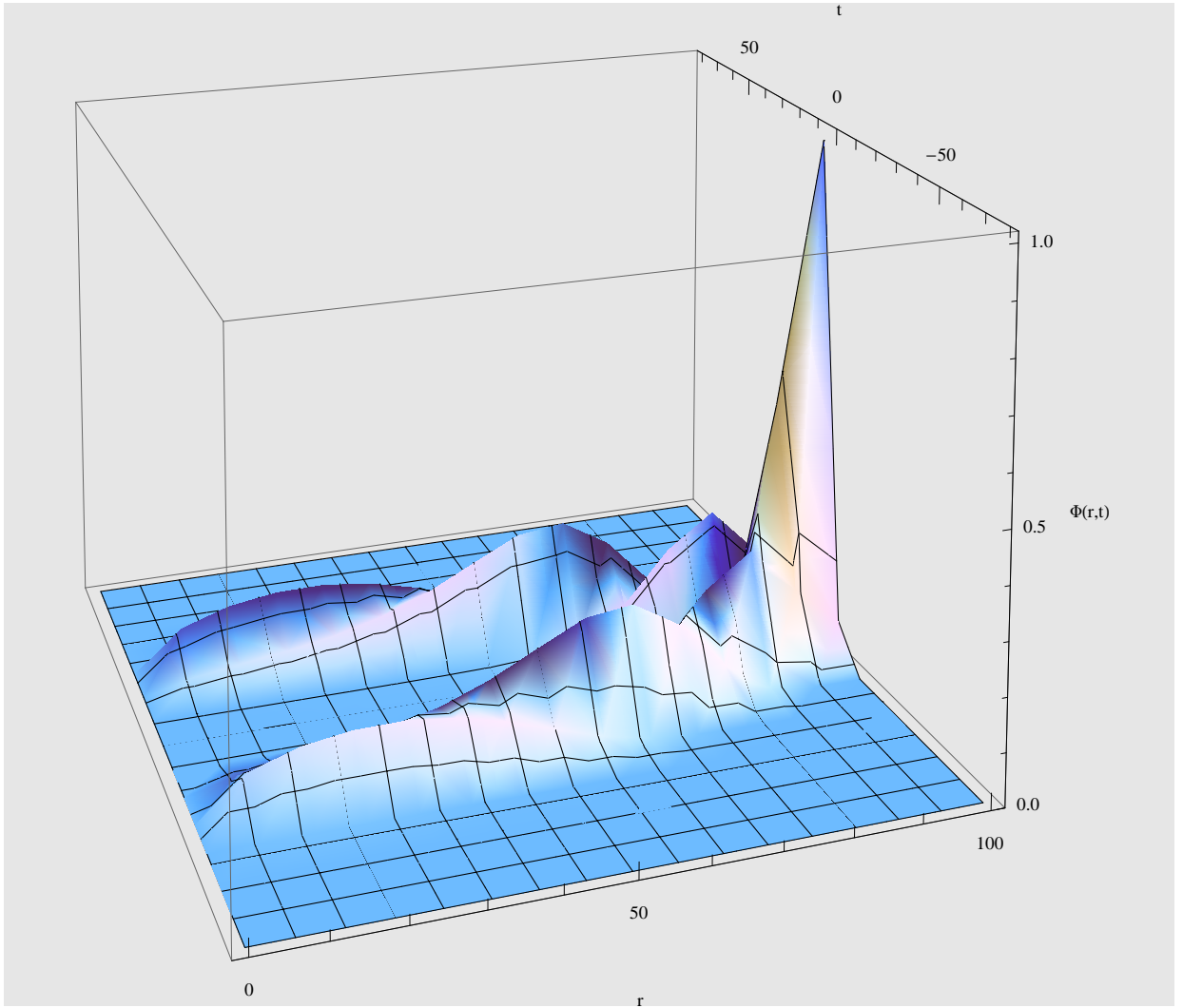}\\
  \caption{The probability density function $\Phi(r,t)\Phi^*(r,t)r^2$ 3D "evolution" from r=M to r=0.}\label{numWaveFunctionM20}
  \end{center}
\end{figure}

After solving numerically equation (\ref{R equation}), we can see in figure \ref{numWaveFunctionM20} the evolution of the wavefunction $\Phi(r,t)$, in the interior region $0<r<M$.

\begin{figure}
\begin{center}
  \includegraphics{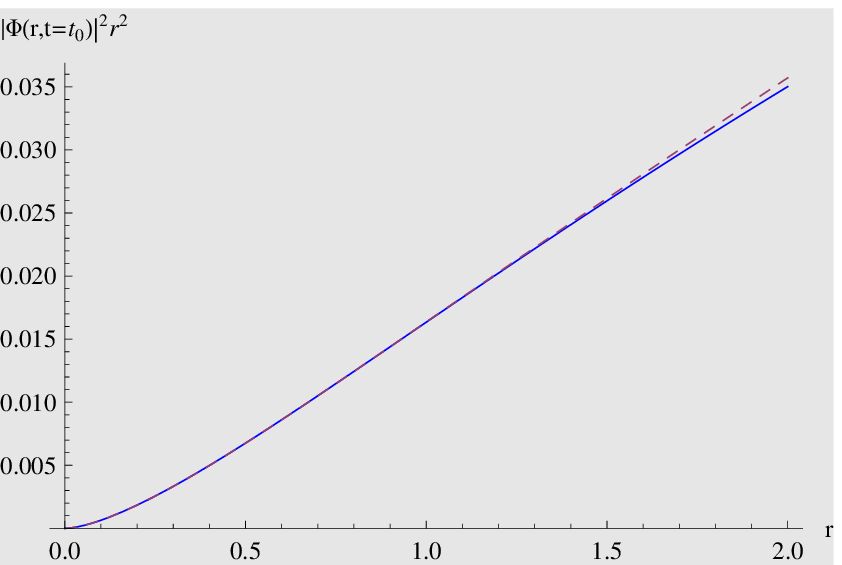}\\
  \caption{The agreement between the analytical to the numerical solution when approaching $r=0$ is shown. The purple dashed line is the analytic solution close to $r=0$. }\label{agreementAtOrigin}
  \end{center}
\end{figure}

Because of the symmetry of the boundary conditions around $t=0$, the function separates into two wave functions, one propagating in the $t$ direction and the other propagating in the $-t$ direction as $r$ decreases.
When approaching to $r=0$, the propagation "freezes" in $t$ and the only thing that changes in the evolution of the wavefunction is the amplitude which decreases as described in our analytic solution according to equation (\ref{eq. wavefunction at origin}). The agreement between the numerical solution and our analytic solution near $r=0$, is shown in figure \ref{agreementAtOrigin}.

\begin{figure}
\begin{center}
  \includegraphics{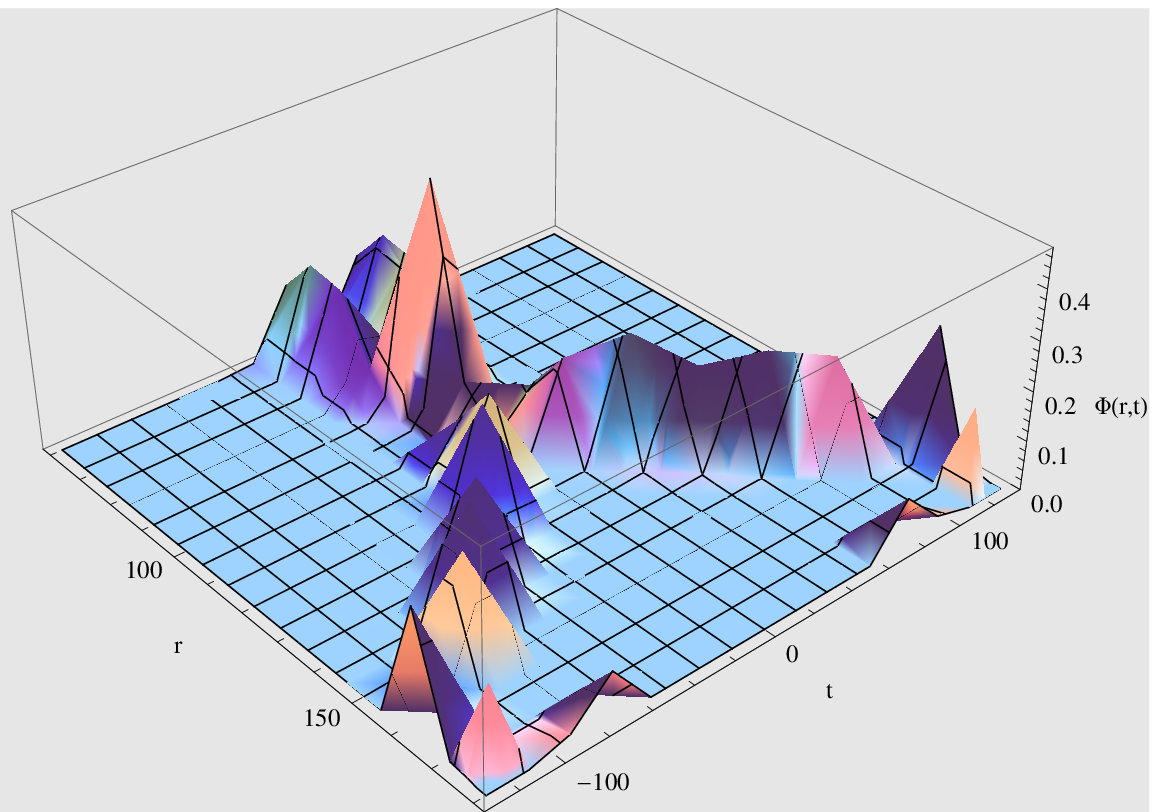}\\
  \caption{The probability density function $\Phi(r,t)\Phi^*(r,t)r^2$ 3D "evolution" from the horizon to r=M.}\label{numWaveFunction2Hor}
  \end{center}
\end{figure}

The evolution of the wavefunction from the horizon to $r=M$ is shown in figure \ref{numWaveFunction2Hor}. It can be seen that the boundary conditions cause two major parts of the wave function at the origin to interfere at $r=M$ and create our gaussian.

\begin{figure}

  \includegraphics{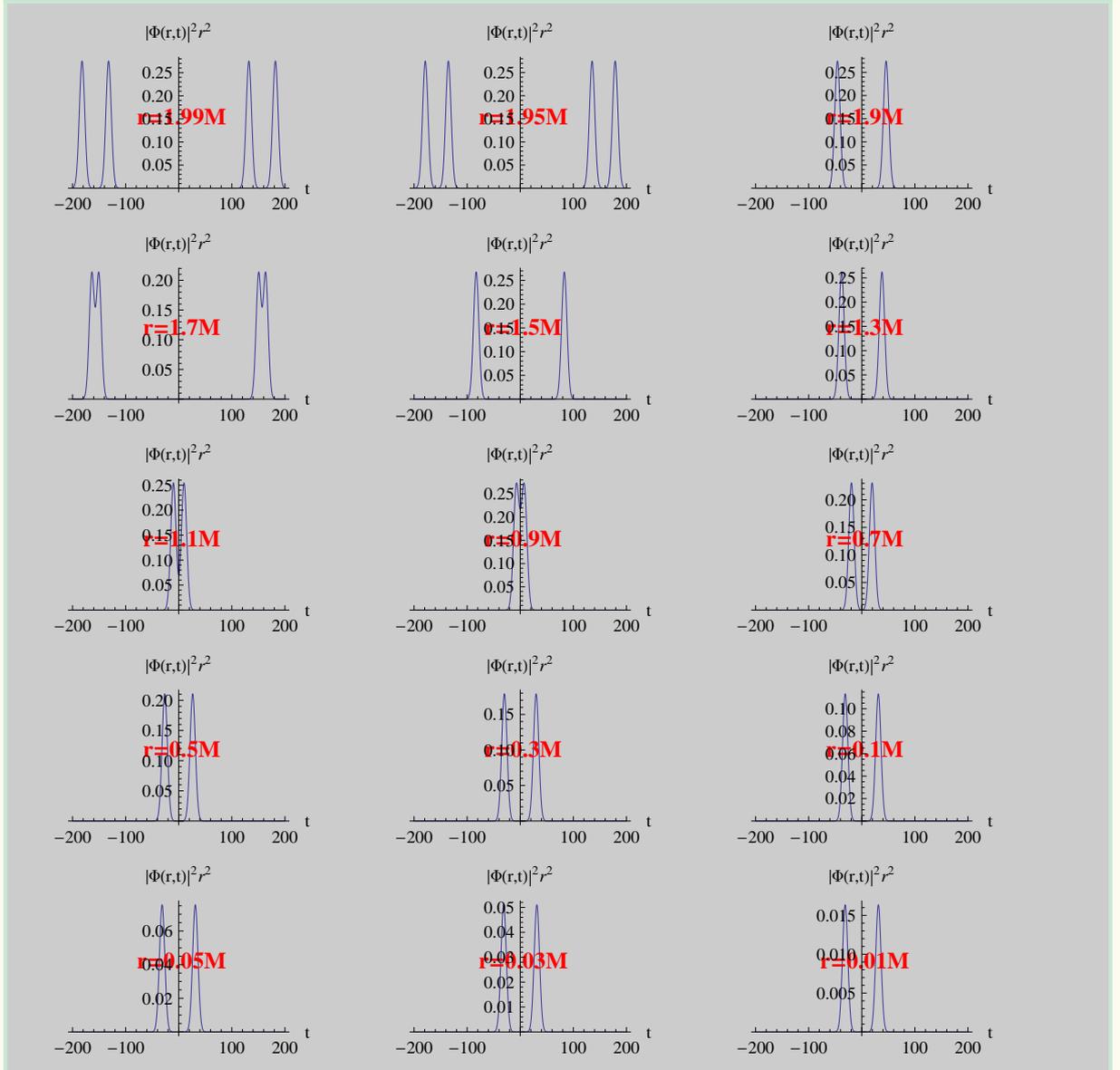}\\
  \caption{$\Phi(r,t)\Phi^*(r,t)r^2$ - 2D cuts from the horizon to $r=0$}\label{matrixCut}

\end{figure}

In figure \ref{matrixCut}, cuts according to $r$ of the 3D probability density function are shown.

The most surprising fact in our solution is the ongoing "disappearing" of the probability density function when approaching $r=0$. According to the classical interpretation it looks as if the particle disappears in it's future, something that is of course impossible. The dependence of the wavefunction in $\tau$ is therefore the only part in the wave function which is "non-localized" and therefore is the only variable that can be interpreted to be the 'time-evolution' of the particle.

Since the numerical wavefunction we have created isn't localized in $r$, it is hard to normalize it around the horizon and therefore, it is impossible to perform an accurate expectation value for $r$. However, integrating over the probability density close to the horizon gives a bigger probability than the integration around $r=M$ as opposed to the wave probability density function at r=0 which is zero. The numerical solution therefore, also strengthens our previous analytic results that the particle is preferably found near the horizon.

\section{Discussion and Summary}

The quantum model we have used in this paper, based on the Stueckelberg-Schrodinger equation, suggests new physics regarding the prediction of a repulsive gravitational behavior within the \emph{black hole}. This result can only be explained quantum mechanically. We have solved the $\tau$ independent equation for $r\rightarrow \infty$, the limit $r\rightarrow 2M^\pm$, and $r \rightarrow 0$, and find that, as expected, the test particle at $2M^+$ becomes trapped, but inside the horizon, it does not fall to the center, but (as for other initial conditions within the horizon) tends to be found around the horizon.
Due to the similarity of the Stueckelberg theory to the structure of the KG (Klein-Gordon) equation, apparently, one could have used the KG equation and get the same results. As a matter of fact, equation (\ref{R equation}) is actually the radial part of the KG equation (if one, considers it as a wave equation).  The KG equation, however, as pointed out by Newton and Wigner \cite{NW49} does not provide a wave function from which one can make local conclusions about the distribution of matter, which is our purpose here. The Stueckelberg theory, however, which is intrinsically off-shell, has the property that it provides a quantum theory with the correct properties of locality \cite{HP73}. It is for this reason that we studied this question in the framework of Stueckelberg. The results of our computations in this framework have a clear interpretation whereas from the point of view of KG theory, they are difficult to interpret.
This is because Stueckelberg's "world time" doesn't necessarily propagate with 'r' within the black hole, while in the KG equation 'r' is the time-like axis where the particle must propagate towards its "future" at $r=0$, and must not "disappear" as is seems to do in a detailed analysis.

We would like to thank Prof. Amos Ori from the Physics department of the Technion Institute for his valuable contributions to this article and especially for his help in constructing a numerical analysis model that corresponds to a correct General Relativity approach.

\end{document}